\documentclass[letter]{aa}  
\newcommand{\jjj}{J1402+5347}
\newcommand{\kms}{km s$^{-1}$}
\newbox\grsign \setbox\grsign=\hbox{$>$} \newdimen\grdimen \grdimen=\ht\grsign
\newbox\simlessbox \newbox\simgreatbox
\setbox\simgreatbox=\hbox{\raise.5ex\hbox{$>$}\llap
     {\lower.5ex\hbox{$\sim$}}}\ht1=\grdimen\dp1=0pt
\setbox\simlessbox=\hbox{\raise.5ex\hbox{$<$}\llap
     {\lower.5ex\hbox{$\sim$}}}\ht2=\grdimen\dp2=0pt

\def\simless{\mathrel{\copy\simlessbox}}

\usepackage{graphicx,upgreek}
\usepackage{txfonts,bm}
\begin{document}

   \title{Extreme intra-hour variability of the radio source \\J1402+5347 discovered with Apertif}
\titlerunning{Extreme intra-hour variability in J1402+5347}
\authorrunning{Oosterloo, Vedantham, Kutkin et al.}

\author{T.A. Oosterloo$^{1,2}$,
H.K. Vedantham$^{1,2}$,
A.M. Kutkin$^{1,3}$,
E.~A.~K.~Adams$^{1,2}$,
B.~Adebahr$^{4}$,
A.H.W.M.~Coolen$^{1}$,
S.~Damstra$^{1}$,
W.J.G.~de~Blok$^{1,2,5}$,
H.~D\'enes$^{1}$,
K.M.~Hess$^{1,2}$,
B.~Hut$^{1}$,
G.M.~Loose$^{1}$,
D.M.~Lucero$^{2,6}$,\\
Y.~Maan$^{1,7}$,
R.~Morganti$^{1,2}$,
V.A.~Moss$^{1,8,9}$,
H.~Mulder$^{1}$,
M.J.~Norden$^{1}$,
A.R.~Offringa$^{1}$,
L.C.~Oostrum$^{1,7}$,\\
E.~Orr\`u$^1$,
M.~Ruiter$^{1}$,
R.~Schulz$^{1}$,
R.H.~van~den~Brink$^{10}$,
J.M.~van~der~Hulst$^{2}$,
J.~van~Leeuwen$^{1,7}$,\\
N.J.~Vermaas$^{1}$,
D.~Vohl$^{1}$,
S.J.~Wijnholds$^{1}$,
J.~Ziemke$^{1,11}$
}

\institute{
$^1$ASTRON, the Netherlands Institute for Radio Astronomy, Oude Hoogeveensedijk 4,7991 PD Dwingeloo, The Netherlands\\
$^2$Kapteyn Astronomical Institute, University of Groningen, Postbus  800, 9700 AV Groningen, The Netherlands \\
$^3$Astro Space Center of Lebedev Physical Institute, Profsoyuznaya Str.\ 84/32, 117997 Moscow, Russia\\
$^4$Astronomisches Institut der Ruhr-Universit{\"a}t Bochum (AIRUB), Universit{\"a}tsstrasse 150, 44780 Bochum, Germany \\
$^5$Department of Astronomy, University of Cape Town, Private Bag X3, Rondebosch 7701, South Africa\\
$^6$Department of Physics, Virginia Polytechnic Institute and State University, 50 West Campus Drive, Blacksburg, VA 24061, USA \\
$^7$Anton Pannekoek Institute, University of Amsterdam, Postbus 94249, 1090 GE Amsterdam, The Netherlands \\
$^8$CSIRO Astronomy and Space Science, Australia Telescope National Facility, PO Box 76, Epping NSW 1710, Australia\\
$^9$Sydney Institute for Astronomy, School of Physics, University of Sydney, Sydney NSW 2006, Australia\\
$^{10}$Tricas Industrial Design \& Engineering, Hanzelaan 95b, 8017 JE Zwolle, The Netherlands\\
$^{11}$Center for Information Technology, University of Groningen, Postbus 11044, 9700 CA Groningen, the Netherlands
}

\date{}

 
  \abstract{
The propagation of radio waves from distant compact radio sources through turbulent interstellar plasma in our  Galaxy causes these sources to twinkle, a phenomenon called interstellar scintillation. Such scintillations are a unique probe of the micro-arcsecond structure of radio sources as well as of the sub-AU-scale structure of the Galactic interstellar medium. 
Weak scintillations (i.e.\  an intensity modulation of a few percent) on timescales of a few days or longer are commonly seen at centimetre wavelengths and are thought to result from the line-of-sight integrated turbulence in the interstellar plasma of the Milky Way. 
So far, only three sources were known that show more extreme variations, with modulations at the level of some dozen percent on timescales shorter than an hour. 
This requires propagation through nearby ($d\simless 10$ pc) anomalously dense ($n_{\rm e}\sim 10^2$   cm$^{-3}$) plasma clouds. Here we report the discovery with Apertif of a source (\jjj) showing extreme ($\sim$50\%) and rapid variations on a timescale of just 6.5 minutes in the decimetre band (1.4 GHz). The spatial scintillation pattern is highly anisotropic, with a semi-minor axis of about 20,000 km. The canonical theory of refractive scintillation constrains the scattering plasma to be within the Oort cloud. The sightline to \jjj,\ however, passes unusually close to the B3 star Alkaid ($\upeta$\,UMa) at a distance of 32\,pc. If the scintillations are associated with Alkaid, then the angular size of \jjj\  along the minor axis of the scintels must be smaller than $\approx 10\,\upmu{\rm as,}$ yielding an apparent brightness temperature for an isotropic source of $\gtrsim 10^{14}$\,K. }


\maketitle
%
\section{Introduction}

Intra-hour variability (IHV)  refers to quasi-random intensity modulation that is similar to canonical interstellar scintillation, but on vastly shorter timescales. It is an extreme manifestation of radio-wave propagation through turbulent plasma, but the nature of the plasma clouds causing the fast variability has remained unclear so far. It is generally thought to be caused by nearby ($d\sim 10$\,pc) turbulent plasma \citep{bignall-2007,debruyn-2015}. Because the scattering strength of plasma is weaker if the plasma clouds are at small distances, the plasma densities inferred from IHV observations are orders of magnitude higher than that in the ambient ionised interstellar gas. The formation and survival of such over-pressured clouds is at the heart of the mystery. No multi-wavelength counterparts to the plasma have yet been found.

A recent  development is the statistical association of three known IHVs with nearby hot stars by \citet{walker-2017}. suggested that the  scattering is caused by the ionised sheath of tiny self-gravitating  molecular clumps similar to the cometary knots seen in the Helix nebula, for example.  If true, this would be extremely interesting as these clouds would constitute a new population of objects within the interstellar medium.  
Alternatively,  \citet{pen-king} and \citet{pen-levin} have refined the model of \citet{goldreich-sridhar}, which suggests that the extreme scattering phenomena are caused by thin plasma sheets (which have a basis in magnetohydrodynamics, MHD, theory) oriented  along the sightline. The implication here being that there is nothing inherently unusual about the intervening plasma save a fortuitous  orientation.

Only very few IHVs are known. More such objects  need to be discovered to better understand their statistical properties and for more secure empirical associations.  More  cases with extreme properties need to be studied to test the limits of proposed theories. The recent spate of wide-field survey radio telescopes operating in the GHz band, such as the Australian Square Kilometre Array Pathfinder (ASKAP) and the Aperture Tile In Focus (Apertif) project on the Westerbork Synthesis Radio Telescope (WSRT), are in an excellent position to provide larger samples. Here  we report on the first IHV discovered by the recently started  survey with the Apertif phased-array frontends on the WSRT.

\section{Discovery and follow-up}

\begin{figure}
    \centering
    \includegraphics[angle=-90,width=\linewidth]{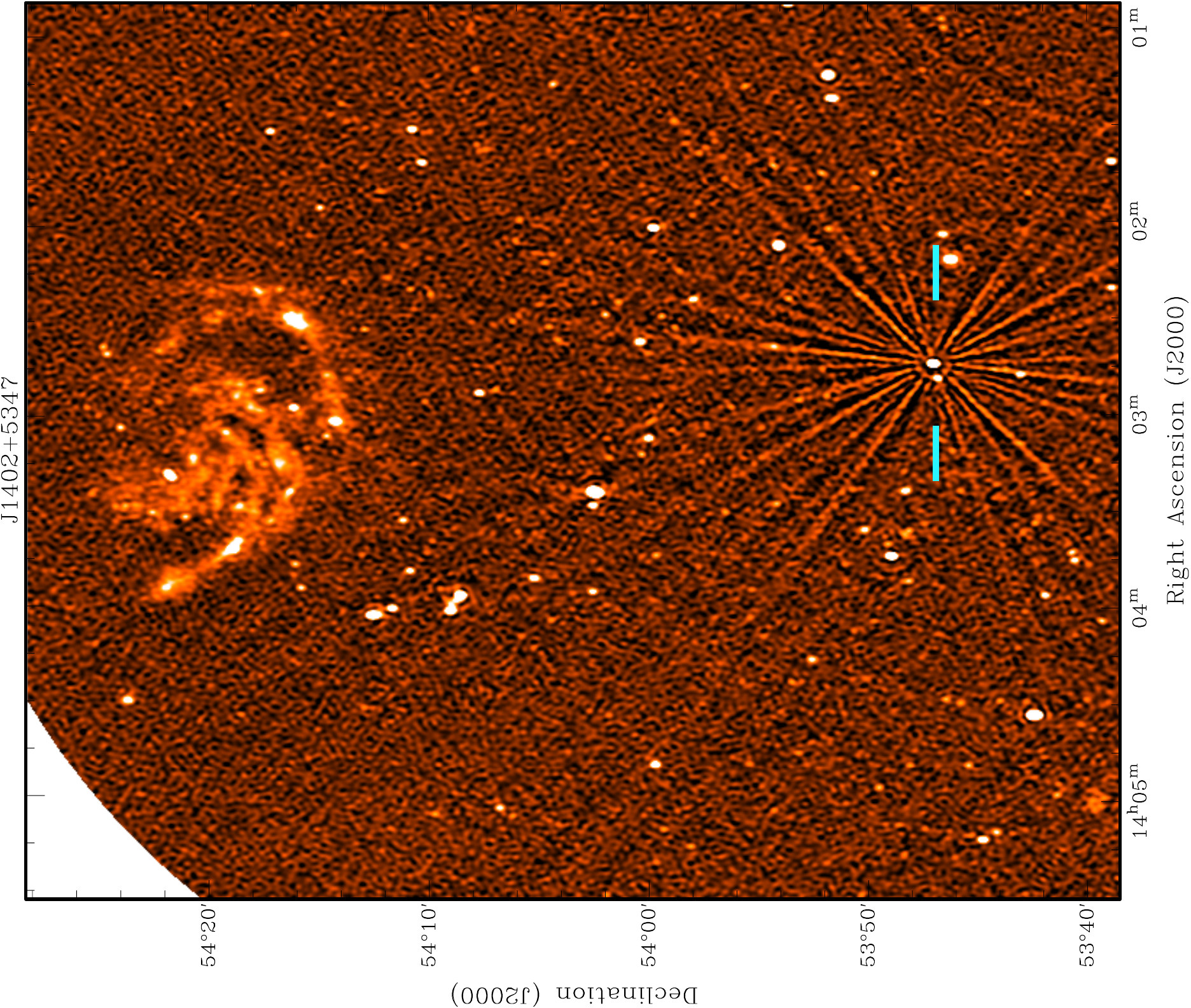}
    \caption{Discovery image of \jjj\ (marked by the blue lines) made from the observation on April 9, 2019. The nearby spiral galaxy M101 lies to the north.  \jjj\ was identified as a highly variable source because of the prominent radial spikes that are centred on it.\label{fig:spikes}  }
\end{figure}

We serendipitously discovered large and rapid scintillations in the radio source \jjj\ (RA$_{\rm J2000}$  14$^{\rm h}$ 02$^{\rm m}$ 43\fs6,  Dec$_{\rm J2000}$ 53\degr 47\arcmin 11\arcsec) with the Apertif phased-array feed system on the WSRT in commissioning data taken on April 9, 2019. Because of the east-west orientation of the  WSRT, highly variable sources show prominent artefacts in the form of radial spikes in long-exposure images, which makes them easy to identify (Fig.\ \ref{fig:spikes}). The light curve  extracted for the source (Fig.\ 2; details in Appendix A)  shows strong flux-density variations that are about $38\%$ of the mean value. Archival WSRT observations from 2013 of the same field at the same frequency showed no significant variability in \jjj. No counterpart to \jjj\ is visible in optical images, but an object is present at
the location of \jjj \ in images produced by the WISE satellite \citep{Cutri2012}. The lack of an optical counterpart together with the  colours of this object in the WISE data suggests that the background object is a dust-obscured active galactic nucleus (AGN) at fairly high redshift ($z\gtrsim 1$; \citealt{Jarrett2017}). 

Follow-up observations taken in May 2019 also showed strong variability at about 30\%. Further observations in June and July 2019 showed weaker and slower variability (Fig.\  2). An annual modulation of the scintillation properties  has been observed in  the other known IHVs \citep{dennett-thorpe-2003,walker-2009,bignall-2019} and is caused by the annual change in the relative velocity of the Earth with respect to the plasma clouds that is due to the orbital motion of the Earth around the Sun. Motivated by this, we continued monitoring the source with a cadence of about one month. The intensity and rate of the variations continued to decrease and \jjj\ virtually ceased to vary in August. Following a brief period of slow variability in September and October, another standstill was observed in November. After November, the rate and intensity of the  variations increased steadily, and by March 2020, they even  exceeded the high levels seen in the discovery observation of April 2019. 
\begin{figure}
    \centering
   \includegraphics[width=\linewidth]{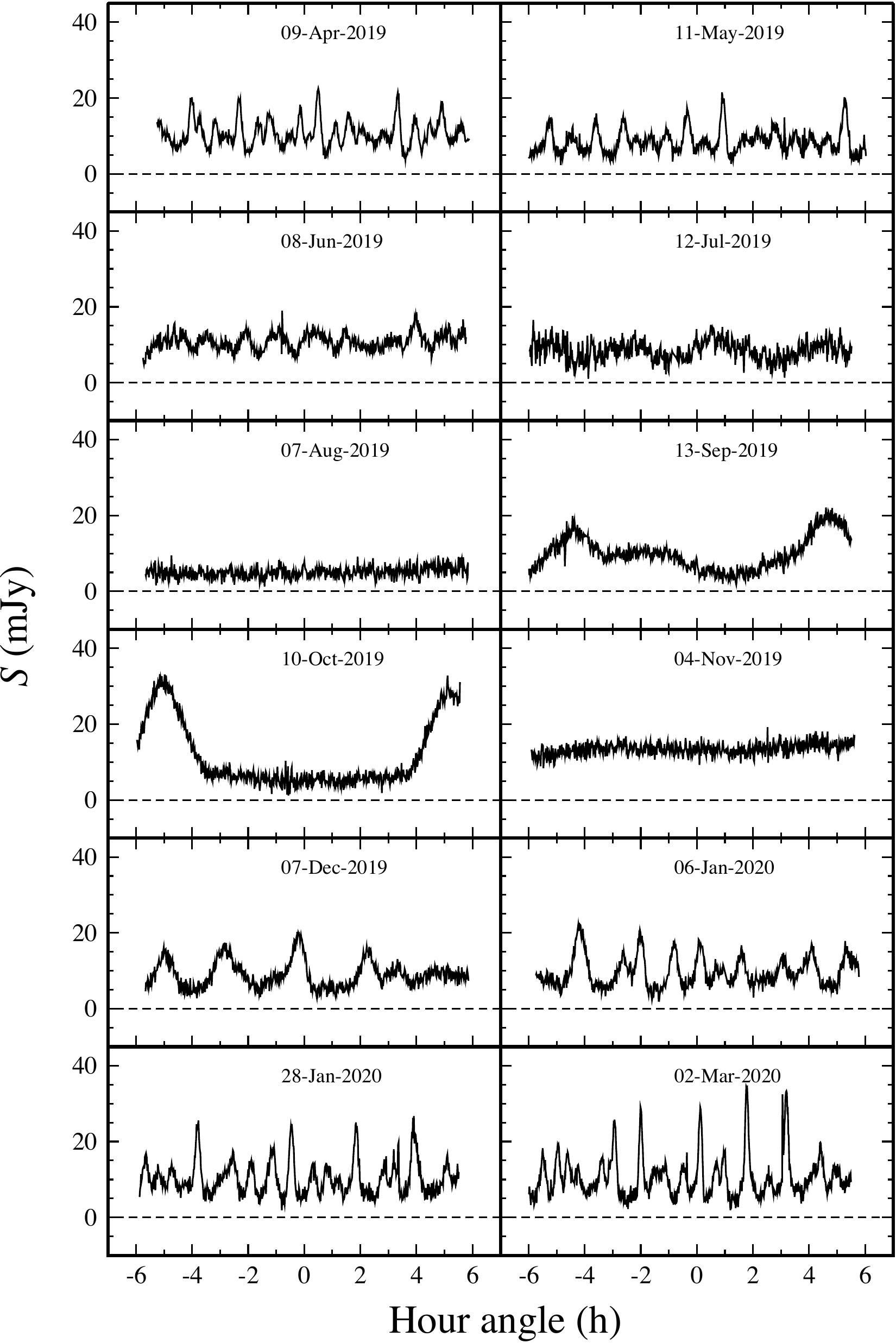}
    \caption{Light curves of \jjj\ observed over a year with a monthly cadence. The observation dates are indicated. The annual modulation in the variation rate is apparent. We observe two standstills in August 2019 and November 2019, while the most rapid variations occurred in April 2019 and March 2020.\label{fig:lc}}
\end{figure}

\section{Timescale analysis}

\begin{figure*}
    \centering
    \includegraphics[width=0.7\linewidth]{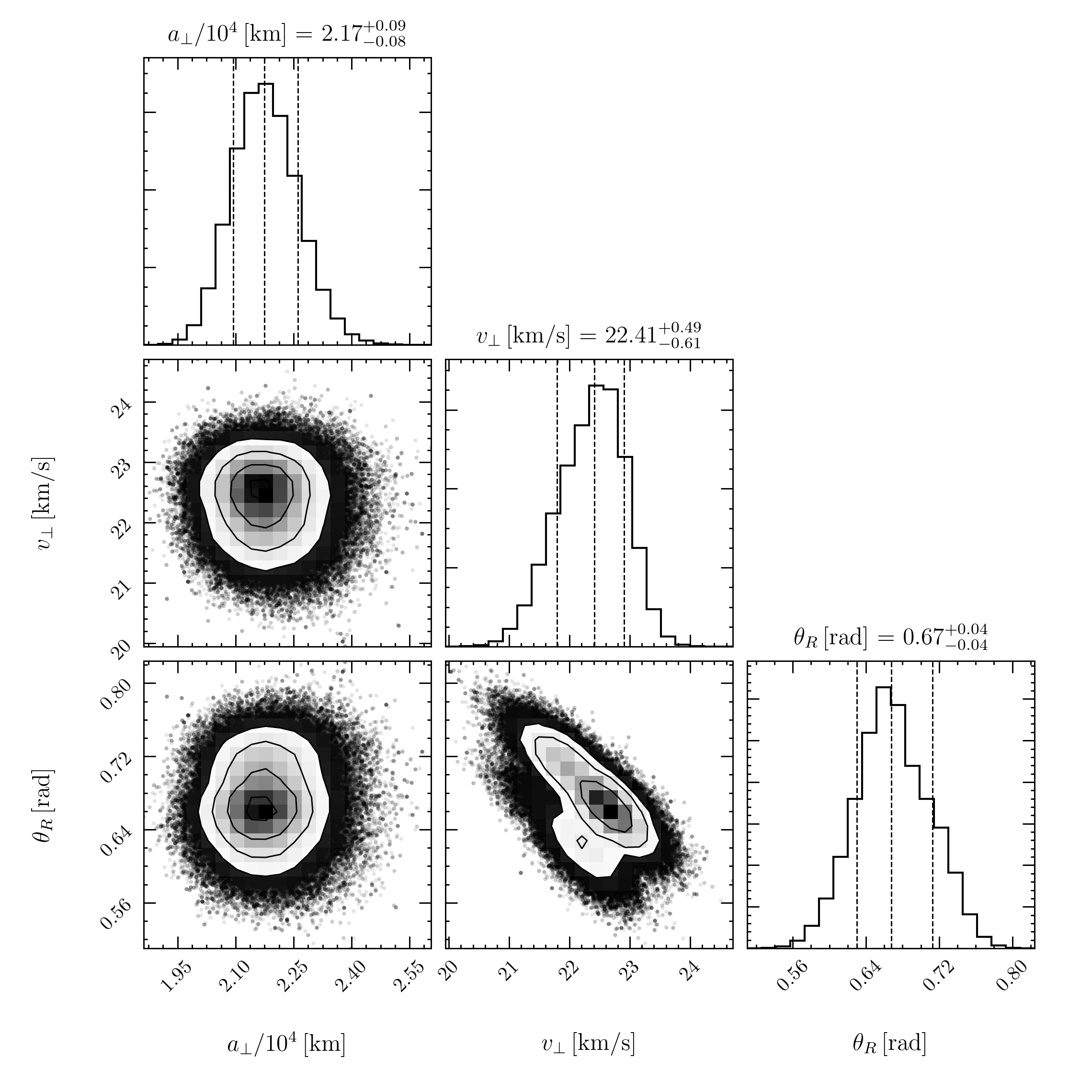}
    \caption{Posterior distribution of the model parameters describing the annual modulation of the scintillation rate. The panels on the diagonal are the marginalised one-dimensional posterior distributions. The dashed black lines are drawn at 0.16, 0.5, and $0.84$ quartiles. The annotated text gives the best-fit parameters and $\pm1\sigma$ bounds implied by the posterior distributions.\label{fig:posterior} }
\end{figure*}

\begin{figure}
    \centering
    \includegraphics[width=0.9\linewidth]{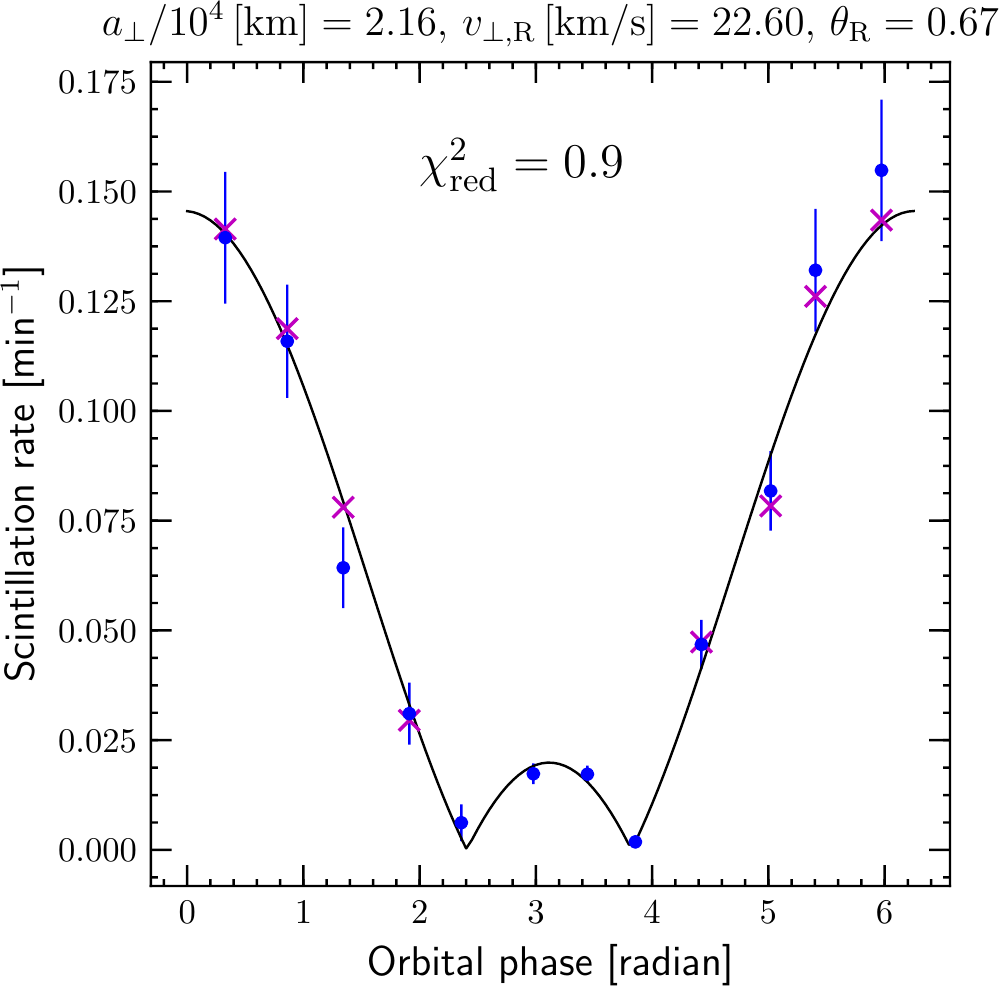}
    \caption{Scintillation rate determined through a Gaussian regression analysis (blue points with $\pm1\sigma$ errors); the maximum likelihood model (black curve) is overplotted. The annotated text gives the model parameters and the reduced $\chi^2$ of the fit. The phase at the vernal equinox is defined to be zero, with phase increasing with time. Magenta crosses show the inverse of the scintillation timescale measured directly from the autocorrelation function (ACF) of the light curves during epochs of sufficiently rapid variations.}
    \label{fig:bestfit}
\end{figure}
The scintillation geometry and physical properties of the plasma clouds causing the scintillation can be determined from the annual variation in the scintillation timescale.
We computed the scintillation timescale for each observation using two methods that are widely used in the  analysis of time-series. The first method determines the decorrelation timescale from the noise-bias-corrected ACF directly computed from the light curves (Appendix B). The temporal lag at which the ACF of the  variations falls below a specified threshold is taken as the scintillation timescale. For easy comparison with previous work \citep{dennett-thorpe-2003,bignall-2019}, we define the threshold as $1/e$ times the peak value. This method is more accurate as it makes minimal assumptions. However, it is only effective for epochs of rapid variability as it requires several scintels to be observed within a single observation run \citep{dennett-thorpe-2003}. For epochs near the standstills, we used a method that is commonly called Gaussian process (GP) regression (Appendix B). This method models the observed variations as a GP with a judiciously chosen functional form for the temporal covariance function. Although this method assumes an analytic form of the covariance function, it can effectively determine the decorrelation timescale and uncertainties even in epochs near the standstills \citep{bignall-2019}. The decorrelation timescales from both methods are tabulated in Table \ref{tab:timescale_data} and agree with each other within the errors for all the epochs to which both methods could be applied.

The scintillation timescale of just $6.5$\,min in March 2020 and the similarly short timescale in the April 2019 observations represent the fastest interstellar scintillation reported to date. This is remarkable because, compared to previous studies, our observations are at lower frequency for which normally the  scintillations  are significantly slower.

This timescale directly relates the physical size of the scintels to the transverse speed between the Earth and the scattering plasma. A single-epoch measurement cannot simultaneously constrain the scintel size and the relative speed. However, the annual modulation in the scintillation timescale can be used to estimate both parameters.
In particular, the two standstills critically constrain both the screen velocity and the geometry of the scintels. One annual standstill can be obtained when the screen velocity on the ecliptic plane equals the Earth's orbital velocity. The second standstill requires the scintels to be anisotropic such that the standstill occurs when the relative velocity between the Earth and the plasma cloud is along the major axis of the scintels. We fit a simple model with   one-dimensional scintels to the observed annual modulation, as suggested by \citet[][see Appendix C]{walker-2009}. The annual variation in this model is fully specified by three parameters: (a) the semi-minor axis of the scintels, $a_{\perp}$, (b) the systemic velocity of the scattering plasma along the minor axis of the scintels, $v_{\perp}$, and (c) the orientation of the major axis on the plane of the sky, $\theta_{\rm R}$ (measured from north to east). Figures \ref{fig:posterior} and \ref{fig:bestfit} show the posterior distributions of the model parameters and  the best-fitting annual modulation curve, respectively. We constrain the semi-minor axis of the scintels to be just $a_\perp = 2.17(0.08)\times 10^4\,{\rm km}$. We note that the reduced chi-squared for the best-fitting model is close to unity, which implies that the one-dimensional scintel model sufficiently and parsimoniously captures the annual variation in the data. To understand the degree of anisotropy implied by the data, we also computed the posterior distribution for a two-dimensional scintillation model (Appendix C). The two-dimensional model places a lower limit on the ratio of the long and short axes of the scintels, $R$, of ${\rm log}_{10}(R) > 2.12$. As expected from the high anisotropy, the data do not constrain the space velocity of the scattering screen parallel to the long axis of the scintels.

\section{Discussion and outlook}

Incoherent synchrotron sources are generally expected to only display refractive interstellar scintillation. Refractive scintels of a point-like source will have a size equal to or larger than that of the first Fresnel zone: $r_{\rm F} = \sqrt{\lambda d/(2\pi)}$, where $\lambda$ is the wavelength and $d$ is the distance to the scattering screen \citep{goodman-2006}. 
If the measured semi-minor axis of the scintels is associated with the Fresnel scale,  the distance to the scattering plasma is $d=0.24\,{\rm pc,}$ which is within the sphere of gravitational influence of the Sun. This would place the scattering plasma within the canonically defined Oort cloud.

If \jjj\ is unusually compact, then it may display diffractive interstellar scintillations, as is usually seen in pulsars. In this case, the scintel size is associated with the diffractive scale. Because the diffractive scale can be much smaller than the Fresnel scale, the data admit significantly larger screen distances in the diffractive regime. Regardless, the length scale of scintels cannot be smaller than the projected size of the source at the scattering screen. We consider for simplicity a circular source of angular radius $\theta_{\rm src}$. The requirement $d\,\theta_{\rm src}<a_\perp$  places a constraint on the apparent brightness temperature of the source of $T_{\rm b}>10^{11}(d/{\rm pc})^2$ {\rm K}.

\begin{figure*}
    \centering
    \begin{tabular}{ll}
    \includegraphics[width=0.4\linewidth]{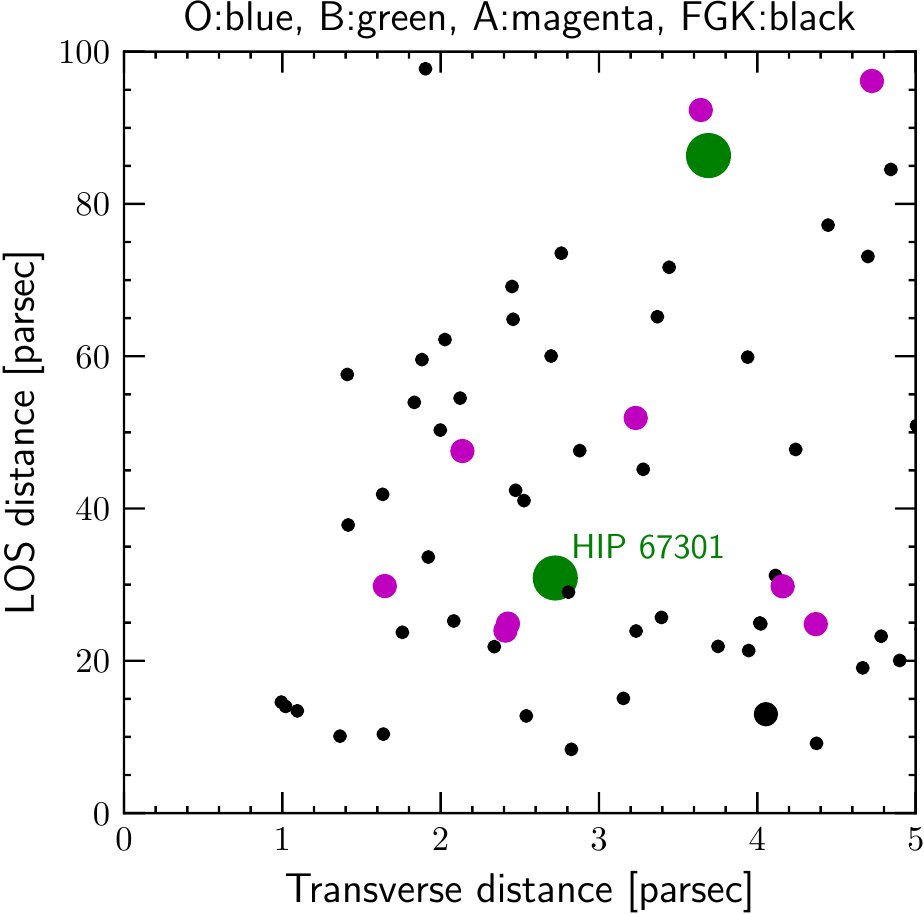} &
    \includegraphics[width=0.4\linewidth]{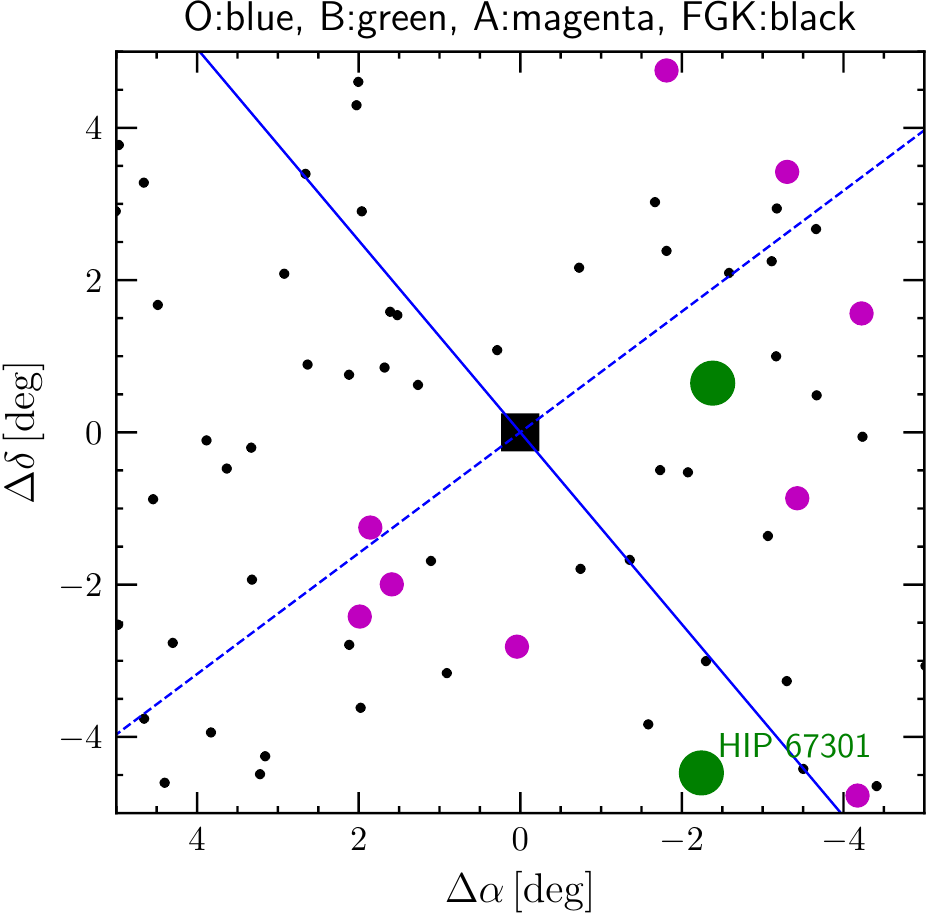}
    \end{tabular}
    \caption{Left panel: Line of sight and transverse distance of stars in the {\it Hipparcos} catalogue with respect to the sightline towards \jjj. The stars are colour-coded by their spectral type (legend in the annotated text). Right panel: Same as the left panel, but the stars are plotted in the plane of the sky. The central black square shows \jjj. The solid and dashed blue lines mark the orientation of the major and minor axes of the scintels, respectively.}
    \label{fig:hip}
\end{figure*}

\citet{walker-2017} have shown a statistical astrometric association between sightlines to sources showing intra-hour variability and hot stars such as Vega, Spica, and Alhakim. They also reported that the long axis of the scintels of IHVs  preferentially points towards the associated star.
We searched the Hipparcos catalogue to verify if the sightline to \jjj\ passes anomalously close to any known O-, B-, or A-type star (see Fig.\ \ref{fig:hip}). The B3 star Alkaid ($\upeta$\,UMa, HIP 67301) at a distance of $d=32$ pc  is unusually close to the sightline towards \jjj. The volume density of B-type stars in the Hipparcos catalogue that lie within a 50-degree-wide cone around this  sightline is about $4.3\times 10^{-5}$ ${\rm pc}^{-3}$. The impact parameter of the sightline with Alkaid is about $2.7$ pc and the angle between the scintel long axis and the star-ward direction for a pairing of \jjj \ and Alkaid is about $9\pm3$\,deg.
The probability of finding a B-type star by chance with the observed impact parameter or closer and with the observed angular alignment or better is about $0.2\%$. In addition, the space velocity components of Alkaid with respect to the Solar System barycentre are $-17.8$ \kms\ and $-2.2$ \kms\ along the RA and DEC axes, respectively. This is about 9 \kms\ different from the space velocity component of the scattering plasma as constrained by our observations, although, as pointed out by \citet{walker-2017}, a clear agreement between space velocities may be precluded because of the free expansion of the scattering plasma, and to a lesser extent, because of projection effects.
Regardless, an association of the scattering plasma with Alkaid sets $d=32$ pc and places a rather extreme lower limit on the apparent brightness temperature of \jjj\ of $T_{\rm b}>10^{14}$ K. This limit can be slightly ameliorated if we postulate that the radio source itself is extended precisely along the major axis of the scintels. 

An association with Alkaid also has critical implications for the variance in the scattering properties of the Galactic plasma at large. The angular broadening expected from the line-of-sight integrated Galactic plasma (unrelated to the scattering screen discussed here)  for this sightline is estimated to be about 0.36 mas according to the widely used NE2001 model \citep{cordes-2002}. The brightness temperature necessary for an association with Alkaid, however, requires that the source diameter does not exceed about 8.6 $\upmu{\rm as}$, which is 40 times smaller than the anticipated angular broadening. An association with Alkaid therefore suggests that the bulk of the scattering in the Galactic interstellar medium must occur in highly localised clumps so that this strong sightline-to-sightline variation can be accounted for.

Whether the scattering plasma is located in the Oort cloud or is associated with Alkaid may be determined
 by extending theoretical analyses, such as the analysis of  \citet{goodman-2006}, to the case of highly anisotropic scintels, and by broadband observations of \jjj\ at higher frequencies. 
If the observed scattering at 1.4 GHz is indeed diffractive,  two effects must be seen in such observations: (a) the scintillations will de-correlate on frequency scales $\Delta\nu/\nu\ll  1$, and (b) the transition to weak scattering must occur at frequencies that are several times higher than 1.4 GHz, and similarly strong light-curve variations must then extend to  these high frequencies. If the scattering is associated with the Oort cloud, then the observed variations are due to refractive scintillations in the weak-scattering regime. We therefore expect to observe broadband variations ($\Delta\nu/\nu\sim 1$) at higher frequencies with a continuously declining level of fractional variations with increasing frequency. The observations of April 9, 2019,  show that the variations at the upper and lower end of the $\Delta\nu/\nu\approx 15\%$ observing band are very similar, with a Pearson correlation coefficient of 0.84 (Appendix D), suggesting refractive variations and hence that the scattering plasma is located in the Oort cloud. However, further theoretical investigation on the spectral decorrelation properties of highly anisotropic scintels are required to clearly determine the distance to the scattering screen.

We end by noting that because of the larger apparent source sizes at lower frequencies, fast intra-hour variability at decimetre wavelengths will be significantly biased towards intervening plasma clouds that are nearby. The discovery of rapid scintillation in \jjj\ in early Apertif data bodes well for ongoing wide-field surveys in the decimetre band. Based on  the discovery of \jjj\ and the sky area covered by the Apertif survey at the time, we expect to discover approximately ten more such systems, and indeed, several new IHV sources have more recently  been identified in Apertif data (Oosterloo et al.\ in prep). 

%
%
\bibliographystyle{aa}
\bibliography{ref}

\begin{table*}
\centering
\caption{Annual modulation statistics: Noise bias (determined from the variance in flux density variation between adjacent points in the light curve) has been subtracted from the modulation index. ACF and GP refer to estimates from direct numerical computation of the ACF and the GP analysis, respectively. 
The errors on the GP scintillation rate estimates do not include the $10\%$ uncertainty on the scintillation rate because of the uncertainty in the kernel parameters (to be added in quadrature). Although scintillation rates were used in the annual modulation fits, we also quote the timescales here for easy comparison to previous work.
ND indicates that not enough scintels are observed for a robust timescale estimation. $1$$\sigma$ uncertainties are given in parentheses. 
In case of significantly asymmetric errors, the $1$$\sigma$ upper and lower bounds are given in the super- and subscript text, respectively.\label{tab:timescale_data}}
\begin{tabular}{lcccc}
{\bf Date}   & {\bf modulation} & {\bf ACF timescale} & {\bf GP timescale} & {\bf GP scintillation}\\
 & {\bf index} & {\bf [min]} & {\bf [min]} & {\bf rate [min$^{\bm -1}$]}\\ 
09-Apr-2019 & 0.31  & $7.1 (0.7)$ & $7.2(0.8)$              &  $0.140(0.006)$    \\ 
11-May-2019 & 0.34  & $8.4(0.9)$  & $8.6(1.0)$              & $0.116(0.006)$       \\
08-Jun-2019 & 0.17  & $13(2)$ & $16(2)$             & $0.064(0.007)$     \\
12-Jul-2019 & 0.22  & $34(8)$  & $32(7)$             &  $0.031(0.006)$   \\
07-Aug-2019 & 0.15  & ND            & $161_{-64}^{+149}$        &  $0.006(0.004)$  \\
13-Sep-2019 & 0.42  & ND            & $57.7_{-5}^{+5.5}$   &  $0.017(0.002)$  \\
10-Oct-2019 & 0.75  & ND            & $58_{-3}^{+3}$   &  $0.017(0.001)$   \\
04-Nov-2019 & 0.07  & ND            & $546_{-193}^{+241}$       &  $0.002(0.001)$    \\
07-Dec-2019 & 0.37  & $21(4)$ & $21(2)$             & $0.047(0.003)$\\
06-Jan-2019 & 0.39  & $13(2)$ & $12(1)$              & $0.082(0.004)$ \\ 
28-Jan-2020 & 0.40  & $7.9(0.9)$  & $7.6(0.8)$               & $0.132(0.005)$  \\
02-Mar-2020 & 0.50  & $7(0.7)$   & $6.5(0.7)$              &  $0.155(0.005)$    \\
\end{tabular}

\end{table*}

\begin{acknowledgements}
We dedicate this paper to the memory of  Ger de Bruyn and of J.P.\ Macquart. HKV thanks Mark Walker for discussions. Credit is also due to Mark Walker for pointing out the possible association with Alkaid.  We thank the anonymous referee for a meticulous reading of the manuscript which  helped us to improve our paper, and for rectifying some errors in notation.
EAKA is supported by the WISE research programme, which is financed by the Netherlands Organisation for Scientific Research (NWO). YM acknowledges funding from the European Research Council under the European Union's Seventh Framework Programme (FP/2007-2013)/ERC Grant Agreement No. 617199. LCO acknowledges funding from the European Research Council under the European Union's Seventh Framework Programme (FP/2007-2013)/ERC Grant Agreement No. 617199.
JMvdH acknowledges funding from the European Research Council under the European Union's Seventh Framework Programme (FP/2007-2013)/ERC Grant Agreement No. 291531 (`HIStoryNU'), JvL acknowledges funding from the European Research Council under the European Union's Seventh Framework Programme (FP/2007-2013) / ERC Grant Agreement No. 617199 (`ALERT'), and from Vici research programme `ARGO' with project number 639.043.815, financed by the Netherlands Organisation for Scientific Research  (NWO).
This work makes use of data from the Apertif system installed at the Westerbork Synthesis Radio Telescope owned by ASTRON. ASTRON, the Netherlands Institute for Radio Astronomy, is an institute of the Dutch Science Organisation (De Nederlandse Organisatie voor Wetenschappelijk Onderzoek, NWO).
\end{acknowledgements}


\begin{appendix}
\section{Interferometric data reduction.}

The observations of \jjj\ presented here were performed with the  Apertif phased-array feed system that was  recently installed on the WSRT. With Apertif, 40 partially overlapping beams, each with a full width at half maximum of about 35 arcminutes, are formed on the sky. The data from each beam are used independently to image the region covered by each beam using standard aperture synthesis. The total area imaged by combining the images from all beams is about $2\overset{\circ}{.}5 \times 2\overset{\circ}{.}5$ with a spatial resolution  of $12^{\prime\prime} \times 17^{\prime\prime}$ at the declination of \jjj.   The observing band was 122 MHz wide, divided into 156 channels, centred on 1.365 GHz. \jjj\ was detected in three overlapping beams.  Each observation had a duration of 11 to 11.5  hours. The calibration of the data from each beam  followed standard procedures, involving excision of radio-frequency interference, flux-, bandpass-, and cross-calibration using 3C147, followed by  self-calibration. The resulting images had small direction-dependent imaging errors, which were corrected for using  standard peeling techniques. This applies in particular to the strong source 3C295, located $2\overset{^\circ}{.}0$ SW of \jjj: because of the large distance from the beam centres, it showed large direction-dependent effects. The resulting images are noise limited with a noise level of about 30~$\mu$Jy beam$^{-1}$ that varies slightly between observations.

To construct the light curves, the source model resulting from the self-calibration was subtracted from the visibilities, except for the source model for \jjj. This created a visibility data set that contained only emission from \jjj. The phase centre of this visibility data set was shifted to the position of \jjj\ so that the phases of all visibilities were zero. Finally, for each time stamp, the visibilities of all baselines and frequencies were summed to obtain the flux density of J1402+5347 for each time stamp. After   primary beam correction, this yields the light curves shown in Fig.\ 2.

\begin{figure*}
    \centering
    \includegraphics[width=0.7\linewidth]{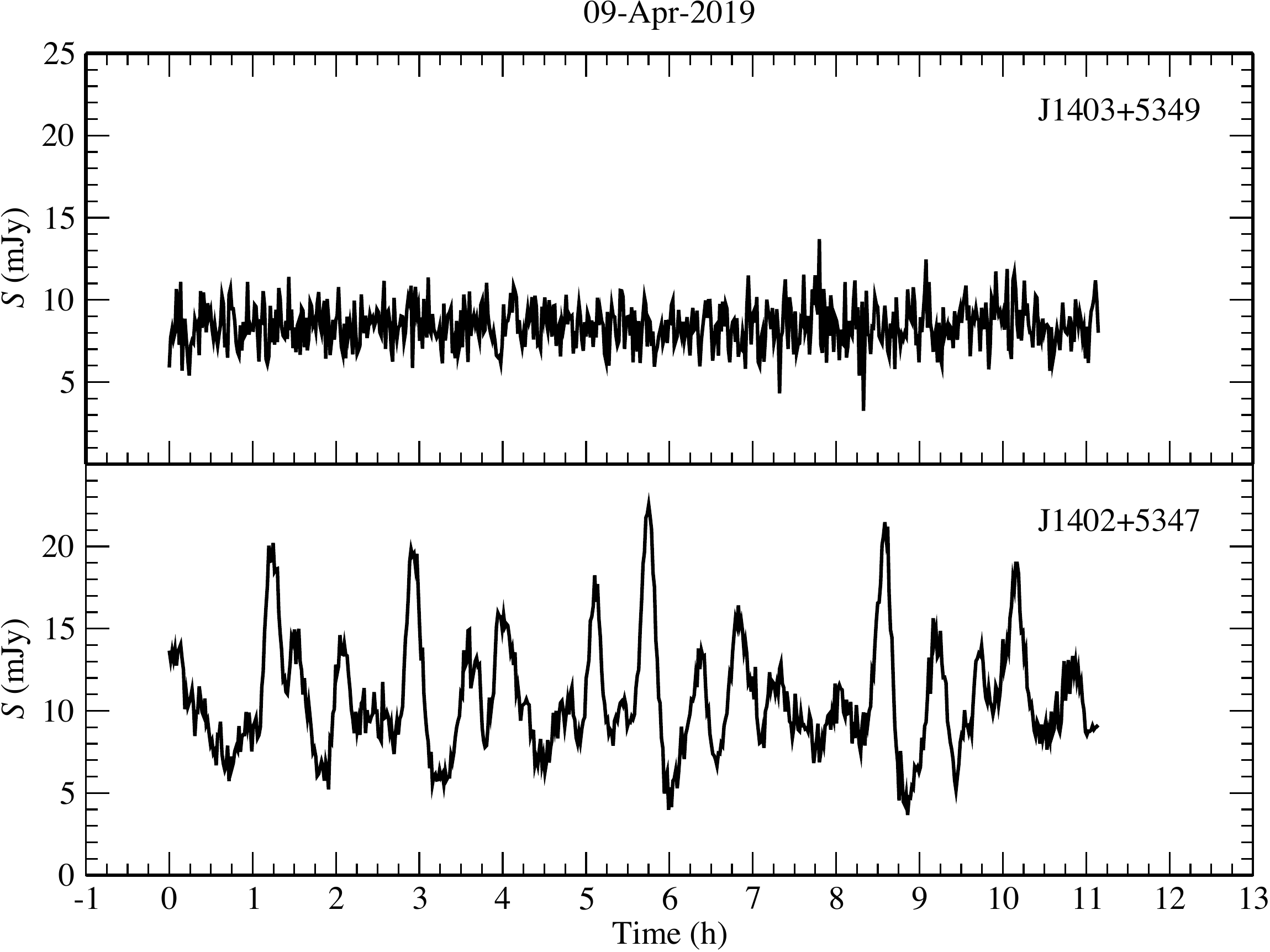}
    \caption{Light curves derived for April 9, 2019, of the variable source \jjj\ and the steady comparison source J1403+5349, 9 arcminutes west of \jjj, illustrating the accuracy of the method for constructing the light curves. The horizontal axis shows the time since the start of the observation.}
    \label{fig:lc_comp}
\end{figure*}

To verify the accuracy of the light curves, the same procedure  was followed for J1403+5349, which has a very similar flux density, 9 arcminutes W of \jjj. An example of such a comparison is shown in Fig. \ref{fig:lc_comp}.

\section{Determination of the scintillation timescale.}
\begin{figure*}
\centering
\includegraphics[width=0.65\linewidth]{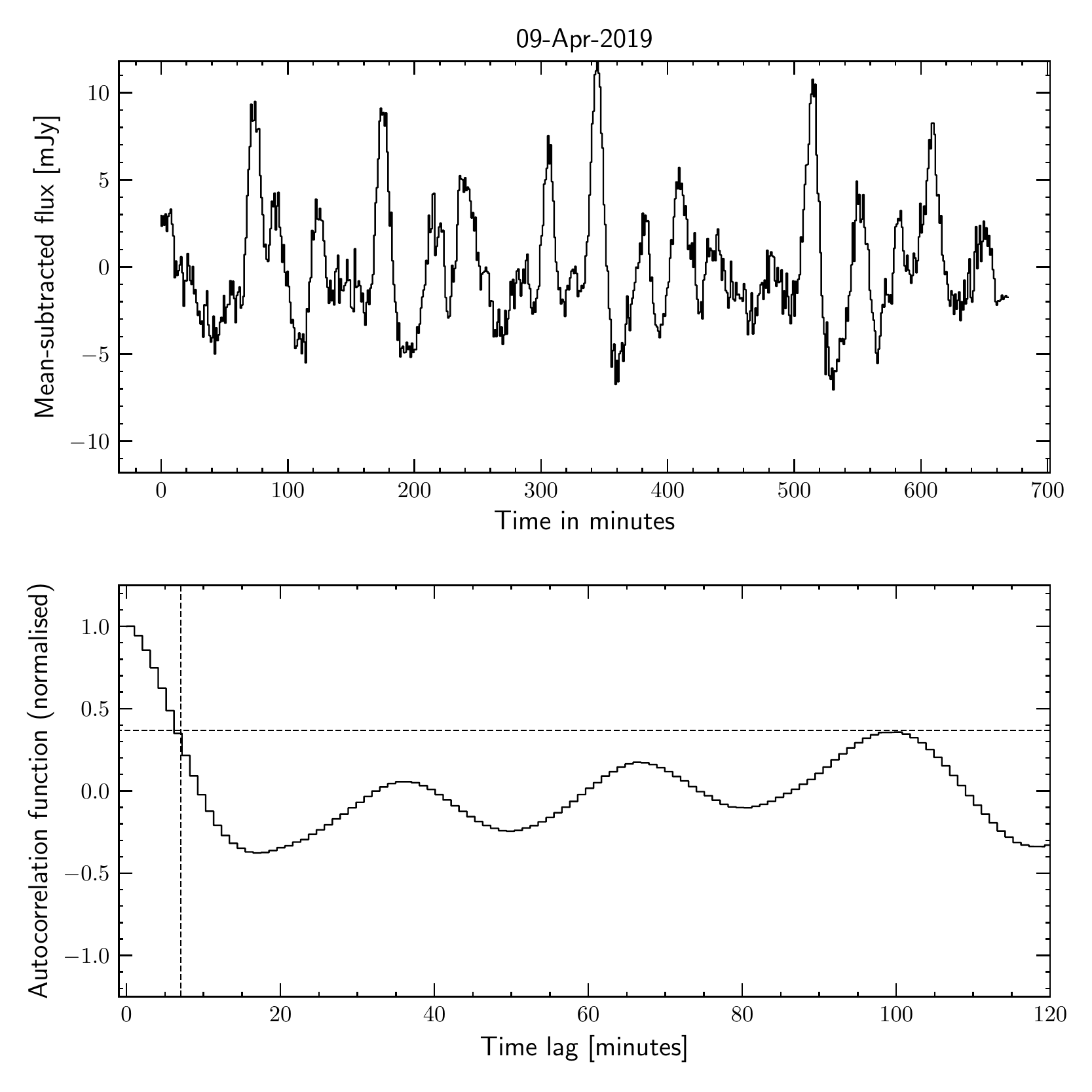}
\caption{Numerical ACF computation: Mean subtracted light curve (top panel) and the ACF (bottom panel) for the April 9, 2019, light curve.\label{fig:acf}}
\end{figure*}

We computed the scintillation timescale using the autocorrelation function (ACF) and Gaussian process modelling (GP).
 For the estimates based on the ACF,  we followed a procedure that closely mimics that presented by  \citet{bignall-2003}. We first  subtract the mean flux density from the light curves to obtain a time series, $x_i$ , with a sampling interval $\Delta t$. For each pair $ij$ we compute the product $x_i x_j$. The list of products is then binned on a grid of size $\Delta t$, and all products falling into the same bin were averaged. Thermal noise only biases the zeroth bin as it is uncorrelated between light-curve samples. This bias, equal to the variance of the noise, is estimated to be equal to half the variance of the difference between successive samples in the time series. The $1/e$ timescale, $t_e$, is estimated using a straight-line fit to a segment of the ACF containing five samples, centred on the sample that is closest to $1/e$. The error on the scintillation timescale is given by $\sigma_{t_e} = \epsilon t_e/\sqrt{N}$, where $N$ is the number of observed cycles of variation on timescale $t_e$ and $\epsilon$ is a factor of about unity that depends on the exact profile of the ACF around $t_e$. If the total duration of observation is $T$ ($\sim$11.5 hr in our case), then $N = (T/t_e)$. The factor $\epsilon$ was estimated with numerical Monte Carlo techniques \citep{dennett-thorpe-2003,bignall-2003} to be in the range $0.25$ to $0.9$. We conservatively assumed $\epsilon = 1$.
    The timescale and its errors thus computed are only accurate when several scintels are observed, which is not the case during the standstills. An example ACF is plotted in Fig.\ \ref{fig:acf}. 
    
\begin{figure*}
\centering
\includegraphics[width=0.75\linewidth]{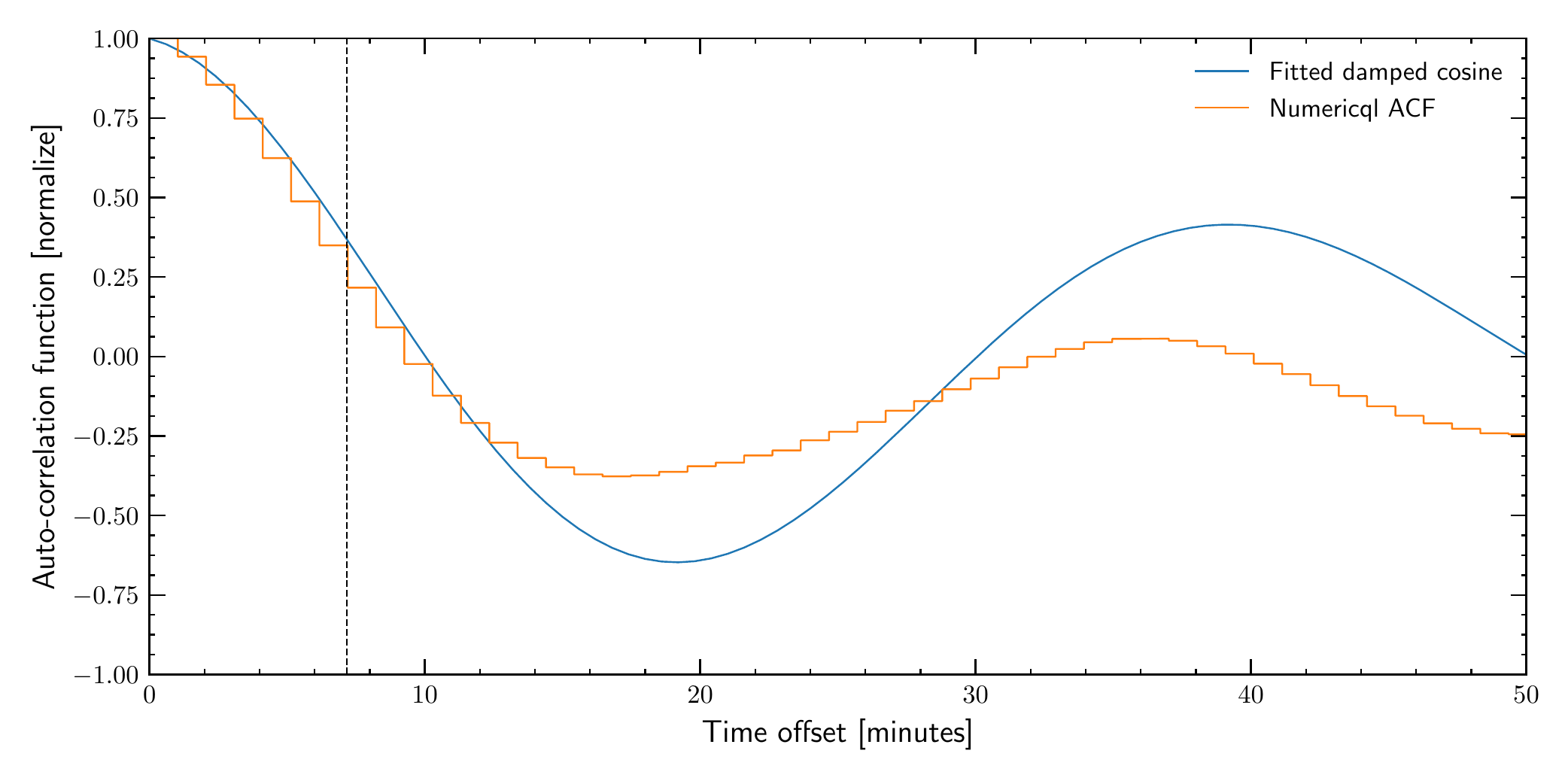}
\caption{Best-fitting kernel: Numerically computed ACF overplotted with the best-fit Gaussian kernel for the  April 9, 2019, light curve. The decorrelation timescale of the Gaussian kernel is indicated with the dashed black line. The quasi-sinusoidal oscillations at long delay are not ideally captured by the analytic form of the kernel, However, the decay to the $1/e$-value at short delays is well captured and is most relevant to estimating scintillation rates. \label{fig:best_fit_kernel} }
\end{figure*}

Our second approach was to model  the temporal scintillation  as a GP with a covariance function given by a damped exponential: $K(\Delta t) = K_0\exp(c\Delta t)\cos(d\Delta t)$, where $\Delta t$ is the temporal lag and $K_0$, $c,$ and $d$ are model parameters \citep{bignall-2019}.  
    This choice for the covariance function is motivated by the shape of the ACF, which shows a steep roll-off at small lags, follows by  oscillatory behaviour due to the dominant spectral mode in the light curves (see Fig.\ \ref{fig:best_fit_kernel})
    We assume that the scintillation is a stationary process, which means that the parameters remain the same at different epochs, and the variation in the scintillation timescale can be modelled by a single epoch-dependent scale factor, $k$, by substituting $\Delta t \rightarrow k\Delta t$. 
    This allows us to estimate the scintillation timescale and the associated uncertainty even for epochs close to the standstill by employing our knowledge of the covariance function from epochs of rapid variability. 
    We determined the parameters $K_0$, $c$, and $d$ from the April 2019 light curve alone instead of using a fit using the entire annual data set \citep{bignall-2019}. With these parameter values fixed, we determined the temporal scale factor $k$ for each epoch  with a single-parameter Gaussian regression. 
    
\begin{figure*}
\centering
\includegraphics[width=0.65\linewidth]{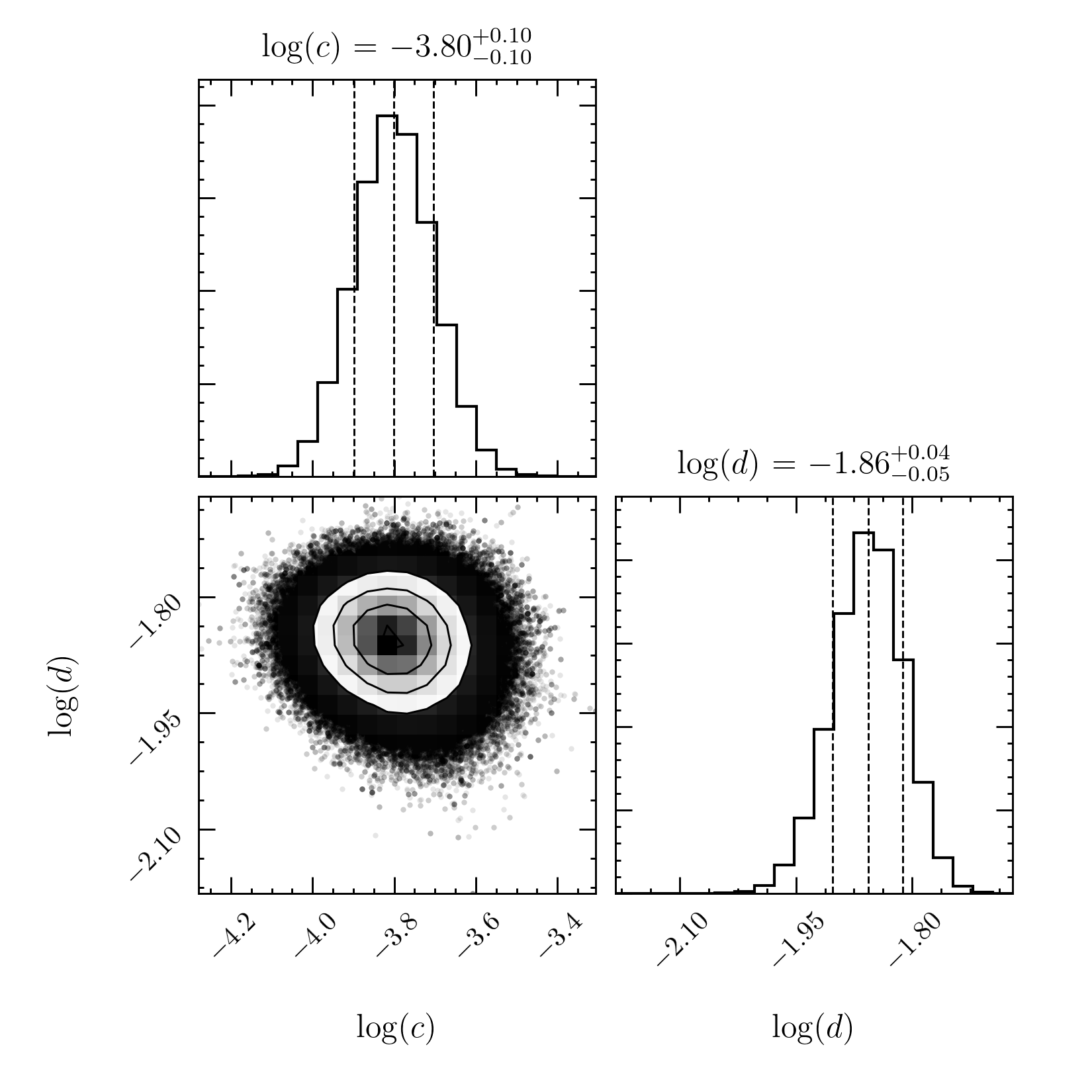}
\caption{Gaussian process kernel parameters: Posterior distribution of GP kernel parameters for the April 9, 2019, data set. The total variance $K_0$ was fixed to equal the noise-bias corrected variance of the light curve, and only the scale parameters $c$ and $d$ were estimated. The dashed lines are placed at the 16, 50, and $84^{\rm th}$ quantiles of the marginalised one-dimensional distributions. The fractional error in the estimation of parameter $c$ is about $10\%$.\label{fig:kernal_params}}
\end{figure*}

    All data were mean subtracted prior to regression analysis, and a noise variance term (estimated from the time differences) was added to the diagonal elements of the covariance matrix. 
    For the April 2019 data set, we fixed $K_0$ to equal the total variance in the scintillations: $K_0 = 3.477105$ ${\rm mJy}^2$, and we estimated $c$ and $d$. The best-fit estimates were $\log(c) = -3.80 \pm 0.10$ and $\log(d) = -1.86 \pm 0.05$. The posterior distributions for these parameters were computed using the \texttt{emcee} software \citep{emcee} and are plotted in Fig. \ref{fig:kernal_params}. The numerically computed ACF and the best-fit kernel are plotted together for comparison in Fig. \ref{fig:best_fit_kernel}.  
 
    Next, we held $K_0$, $c$ and $d$ fixed at the best-fit values quoted above and used the  \texttt{celerite} \citep{celerite} package to fit a single scale parameter $k$.
    The $1/e$ crossing timescale for the April 2019 epoch is largely set by the value of parameter $c$, which is determined with a formal error of about $10\%$. To account for this uncertainty in our error budget, we added a $10\%$ error in quadrature to the formal errors on the timescale from the GP regression analysis. 
    
    Throughout our analysis, the likelihood computations were made using routines from the \texttt{celerite} package \citep{celerite}, the best-fit values were determined using \texttt{scipy.optimize} with the \texttt{L-BFGS-B} method, and the posterior distributions were sampled using the \texttt{emcee} package \citep{emcee}.

\section{Annual modulation.}
The timescale of a temporal scintillation caused by anisotropic scattering is given by \citep{walker-2009}
\begin{equation}
    \label{eqn:ts}
    t_{\rm s} = \frac{R\,a_\perp}{\sqrt{v^2_{||\,{\rm rel}}+R^2v_{\perp\,{\rm rel}}^2}}
,\end{equation}
where $R = a_{||}/a_\perp$ is the anisotropy of the scintels, $a_\perp$ and $a_{||}$ are the length scales of scintels  along the minor and major axes, respectively, and $v_{\perp\,{\rm rel}}$ and $v_{||\,{\rm rel}}$ are the relative velocity between the Earth and the scintels along the minor and major axes of the scintels, respectively. For highly anisotropic scintels \citep{walker-2009}, we can set $R\rightarrow \infty$ and obtain $t_{\rm s} = a_\perp/v_{\perp\,{\rm rel}}$. The annual modulation profile can be fit with a three-parameter model. The parameters are (i) the semi-minor axis of the scintels, $a_\perp$, (ii) the orientation of the major axis of the scintels in the plane of the sky, $\theta_{\rm R}\in(0,2\pi]$,  measured from north towards east, and (iii) the screen velocity in the plane of the sky perpendicular to the major axis of the scintels $v_{\perp}\ge 0$ (defined to be along $\theta_R-\pi/2$).
We carried out our fits in a tangent plane perpendicular to the sightline with cardinal axes along the directions of increasing RA ($\bm{\hat{\alpha}}$) and DEC ($\bm{\hat{\delta}}$).
If ${\bm v_\oplus}(t) = \bm{ \hat{\alpha}}v_{\oplus}^\alpha + \bm{ \hat{\delta}}v_\oplus^\delta $ is the velocity of the Earth projected onto this plane, then the relative velocity perpendicular to the major axis of the scintels is given by $v_{\perp\,{\rm rel}} = |v_{\perp} -v_\oplus^\delta\sin\theta_{\rm R} + v_\oplus^\alpha\cos\theta_{\rm R}|$.
We fit this equation to the scintillation timescales tabulated in Table \ref{tab:timescale_data} (column  ``GP scintillation rate''). We carried out the fits to the scintillation rate instead of the scintillation timescale to avoid infinities in the model. 
First, we carried out a brute-force search for the maximum likelihood value on a coarse three-dimensional parameter grid spanned by $10^8$ cm $<a_\perp<5\times 10^{10}$ cm, $0\,\,{\rm km\,s}^{-1}  <v_{\perp}<100\,\,{\rm km\,s}^{-1}$, and $0<\theta_{\rm R}<2\pi$.  The likelihood peaked at about $a_\perp\approx 2\times 10^9{\rm cm}$, $v_{\perp}\approx 22\,{\rm km\,s}^{-1}$ , and $\theta_{\rm R}\approx 0.7$. We then used the \texttt{emcee} software \citep{emcee} to obtain the final fit values and their formal errors. We used 200 randomly initialised walkers, each walking 2000 steps. We imposed a flat prior on the parameters in the range $2\times 10^8\,{\rm cm}<a_\perp<2\times 10^{10}\,{\rm cm}$, $15\,{\rm km\,s}^{-1}<v_{\perp,\,{\rm R}}<35\,{\rm km\,s}^{-1}$, and $0.1<\theta_{\rm R}<1.1$. The posterior parameter distributions were computed and plotted from the \texttt{emcee} samples using the \texttt{corner} package.

\begin{figure*}
\centering
\includegraphics[width=\linewidth]{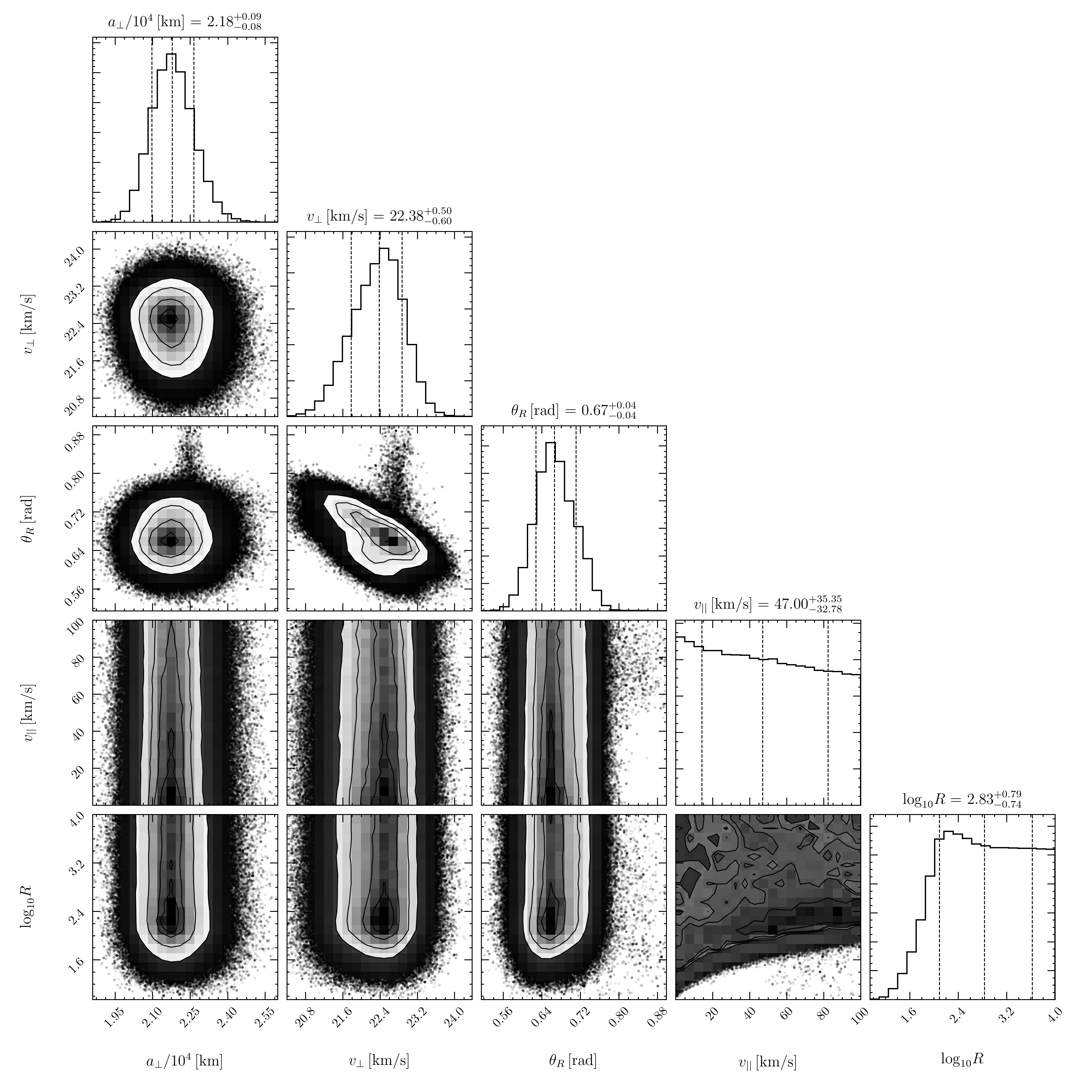}
\caption{Same as Fig. \ref{fig:posterior}, but for a two-dimensional scintel model given in equation \ref{eqn:ts}.\label{fig:2dpcorner}}
\end{figure*}

We also ran a separate MCMC simulation to sample the posterior distribution for the two-dimensional scintillation model given in equation \ref{eqn:ts}. The additional parameters in the two-dimensional model and $v_{||}$ and $R$. We placed the prior bounds $v_{||}\in[0,100)\,{\rm km/s}$ and ${\rm log}_{10}R\in[0,4)$, with the relative velocity parallel to the scintel long axis given by $v_{||\,{\rm rel}}  = \left|v_{||} -v_\oplus^\delta\cos\theta_{\rm R} - v_\oplus^\alpha\sin\theta_{\rm R}\right|$. The posterior distributions are given in Fig. \ref{fig:2dpcorner}.

\section{Broad-band nature of scintels}

\begin{figure*}
\centering
\includegraphics[width=0.75\linewidth]{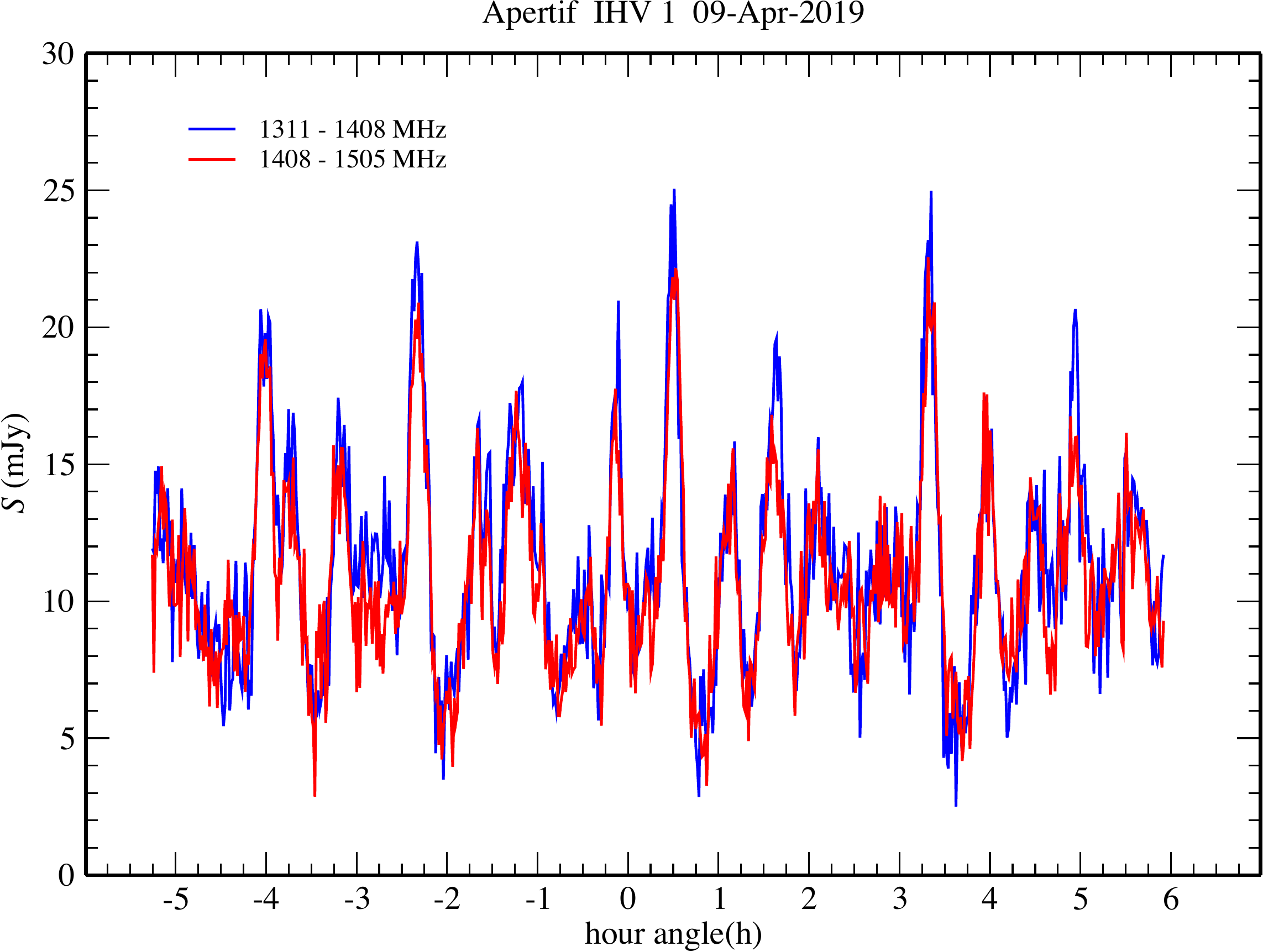}
\caption{Separate light curves  for the lower and upper halves of the observing band of the discovery observation on April 9, 2019. The blue line is the light curve for the frequency range 1311 -- 1408 MHz, and the red line shows the range 1408 -- 1505 MHz.\label{fig:halves}}
\end{figure*}

To investigate a possible frequency dependence of the intensity fluctuations, we constructed separate light curves for the upper and lower halves of the observing band  for the observations on April 9, 2019. These separate light curves are shown in Fig.\ \ref{fig:halves}, where the blue line represents the light curve for the frequency range 1311 -- 1408 MHz and the red line shows the range 1408 -- 1505 MHz. It is clear from this figure that the light curves are very similar, showing that there is no frequency structure with the observing band. The Pearson correlation coefficient between the two light curves is 0.84.

\end{appendix}
\end{document}